\documentclass[10pt,superscriptaddress,aps,prx,twocolumn,nofootinbib]{revtex4-2}
\usepackage{amsmath,amssymb,amsthm,amsfonts}
\usepackage[breaklinks=true,colorlinks,citecolor=blue,linkcolor=blue,urlcolor=blue]{hyperref}
\usepackage[pdftex]{graphicx}
\usepackage{txfonts}
\usepackage{braket}
\usepackage{comment}
\usepackage{siunitx}
\usepackage{float}
\usepackage{algorithm}
\usepackage{algpseudocode}
\usepackage{mathtools}
\usepackage[normalem]{ulem}
\usepackage{tabularx, booktabs}
\usepackage{graphicx}
\usepackage{amsfonts}
\usepackage{color,soul}

\usepackage{url}
\usepackage{orcidlink}

\newcommand{\equalcontrib}{\thanks{These authors contributed equally to this work.}}
\newcommand{\eqlabel}[1]{Eq.~\eqref{#1}}

\newcommand{\tablabel}[1]{Table~\ref{#1}}

\begin{document}

\title{Runtime Quantum Advantage with Digital Quantum Optimization}

\author{Pranav Chandarana$^{\orcidlink{0000-0002-3890-1862}}$}
\equalcontrib
\affiliation{Kipu Quantum GmbH, Greifswalderstrasse 212, 10405 Berlin, Germany}
\affiliation{Department of Physical Chemistry, University of the Basque Country UPV/EHU, Apartado 644, 48080 Bilbao, Spain}

\author{Alejandro Gomez Cadavid$^{\orcidlink{0000-0003-3271-4684}}$}
\equalcontrib
\affiliation{Kipu Quantum GmbH, Greifswalderstrasse 212, 10405 Berlin, Germany}
\affiliation{Department of Physical Chemistry, University of the Basque Country UPV/EHU, Apartado 644, 48080 Bilbao, Spain}

\author{Sebastián V. Romero$^{\orcidlink{0000-0002-4675-4452}}$}
\equalcontrib
\affiliation{Kipu Quantum GmbH, Greifswalderstrasse 212, 10405 Berlin, Germany}
\affiliation{Department of Physical Chemistry, University of the Basque Country UPV/EHU, Apartado 644, 48080 Bilbao, Spain}

\author{Anton Simen$^{\orcidlink{0000-0001-8863-4806}}$}
\equalcontrib
\affiliation{Kipu Quantum GmbH, Greifswalderstrasse 212, 10405 Berlin, Germany}
\affiliation{Department of Physical Chemistry, University of the Basque Country UPV/EHU, Apartado 644, 48080 Bilbao, Spain}

\author{Enrique Solano$^{\orcidlink{0000-0002-8602-1181}}$}
\email{enr.solano@gmail.com}
\affiliation{Kipu Quantum GmbH, Greifswalderstrasse 212, 10405 Berlin, Germany}

\author{Narendra N. Hegade$^{\orcidlink{0000-0002-9673-2833}}$}
\email{narendrahegade5@gmail.com}
\affiliation{Kipu Quantum GmbH, Greifswalderstrasse 212, 10405 Berlin, Germany}

\date{\today}
\begin{abstract}
We demonstrate experimentally that the bias-field digitized counterdiabatic quantum optimization (BF-DCQO) algorithm on IBM’s 156-qubit devices can outperform simulated annealing~(SA) and CPLEX in time-to-approximate solutions for specific higher-order unconstrained binary optimization (HUBO) problems. We suitably select problem instances that are challenging for classical methods, running in fractions of minutes even with multicore processors. On the other hand, our counterdiabatic quantum algorithms obtain similar or better results in at most a few seconds on quantum hardware, achieving runtime quantum advantage. Our analysis reveals that the performance improvement becomes increasingly evident as the system size grows. Given the rapid progress in quantum hardware, we expect that this improvement will become even more pronounced, potentially leading to a quantum advantage of several orders of magnitude. Our results indicate that available digital quantum processors, when combined with specific-purpose quantum algorithms, exhibit a runtime quantum advantage even in the absence of quantum error correction.
\end{abstract}

\maketitle

\section{Introduction}

Combinatorial optimization problems arise in diverse domains such as logistics, manufacturing, finance and chemistry, where finding optimal or near‐optimal solutions is computationally intensive. In the worst case, the runtime on classical hardware grows exponentially with problem size. These problems are commonly tackled with both domain‐specific solvers and general‐purpose algorithms such as simulated annealing (SA)~\cite{kirkpatrick1983optimization}, CPLEX~\cite{cplex}, and Gurobi~\cite{gurobi}. However, these methods struggle with increasingly complex instances. Quantum computing has therefore emerged as a promising alternative for advancing the state of the art in optimization. Early connections between combinatorial optimization and disordered systems such as spin glasses~\cite{fu1986application} inspired the application of statistical physics techniques, particularly the mapping of optimal solutions to ground states of Ising models~\cite{lucas2014ising}. Building on this framework, quantum algorithms such as adiabatic quantum optimization (AQO)~\cite{albash2018adiabatic} and the quantum approximate optimization algorithm (QAOA)~\cite{farhi2014quantumapproximateoptimizationalgorithm}, provide new approaches to these demanding problems~\cite{abbas2024challenges}.

As quantum hardware improves, recent experiments suggest that quantum devices may soon outperform classical solvers on specific families of optimization problems. A variety of quantum‐enhanced strategies have been proposed to accelerate optimization, providing both theoretical speedups and encouraging empirical results~\cite{durr1999quantumalgorithmfindingminimum,montanaro2018quantum-walk,montanaro2020quantumspeedup,chakrabarti2022universalquantumspeedupbranchandbound,somma2008quantum,wocjan2008speedup,hastings2018shortpathquantum,dalzell2023mind}.

One particularly complex class of problems is higher-order unconstrained binary optimization (HUBO). Proof-of-concept implementations on current quantum devices have used AQO and QAOA~\cite{pelofske2023quantum, pelofske2024short-depth, barron2023provableboundsnoisefreeexpectation, sachdeva2024quantum,mcgeoch2024comment, drăgoi2025approximatequadratizationhighorderhamiltonians, doi:10.1126/sciadv.adm6761,10313612}. However, high noise levels in these devices still limit scalability. To overcome this challenge, researchers have increasingly adopted counterdiabatic (CD) driving~\cite{PhysRevApplied.15.024038,demirplak2003adiabatic, berry2009transitionless, chen2010fast, campo2013shortcuts, chandarana2022digitized, hegade2022digitized}. CD techniques aim to suppress non-adiabatic transitions, enabling more accurate and efficient quantum evolution toward optimal solutions.

In this work, we experimentally demonstrate that the recently proposed bias-field digitized counterdiabatic quantum optimization (BF-DCQO) algorithm~\cite{cadavid2024bias, romero2024bias, iskay}, when executed on IBM quantum hardware~\cite{ibm}, can potentially yield faster approximate solutions than classical solvers for specific problem classes. Given hardware constraints, including restricted qubit connectivity and finite coherence times, we develop an instance-generation strategy that enables efficient implementation of relatively dense HUBO instances on current quantum hardware. Guided by our experimental findings, we identify signatures of a runtime advantage using BF-DCQO. Particularly, we benchmark against SA and CPLEX using up to 48 cores and 10 threads, respectively.

The remainder of this article is organized as follows. Section~\ref{sec:problemclasses} describes the procedure for generating challenging HUBO instances and classifies them into two distinct types. Section~\ref{sec:solvers} details the benchmarking methodology for evaluating solver performance. Section~\ref{sec:results} presents the experimental implementation of BF-DCQO on IBM hardware and analyzes its performance relative to classical solvers. Finally, Section~\ref{sec:conclusion} summarizes our conclusions, and the Appendices provide additional remarks and extended results.

\section{Higher-order binary optimization}\label{sec:problemclasses}

Many academic and industrial optimization problems can be formulated as HUBO problems, characterized by a cost function of the form
\begin{equation}\label{eq:binary_hubo}
F(x) = \sum_{k=1}^{d} \sum_{(i_1,\dots,i_k)\in G_{k}}
T_{i_1\cdots i_k}\,x_{i_1}\cdots x_{i_k},
\end{equation}
where $x_i\in\{0,1\}$, $d$ is the maximum interaction order, and the coupling coefficients $T_{i_1\cdots i_k}$ are sampled from an appropriate probability distribution. Here, $G=\bigl(V,\{G_k\}_{k=1}^d\bigr)$ is a hypergraph with $V$ the set of vertices and $G_k$ the set of $k$-vertex hyperedges. The complexity of a HUBO instance is governed by four factors: the number of variables $N$, the interaction coefficients $T_{i_1\cdots i_k}$~\cite{kalinin2022computational}, the maximum interaction order $d$, and the density imposed by the hypergraph connectivity.

We consider systems up to $N=156$ qubits on IBM’s superconducting quantum processors~\cite{ibm}. We restrict $d=3$ to facilitate the experimental demonstrations. After mapping each binary variable $x_i$ to a spin variable via $x_i = \tfrac{1-\sigma_i^z}{2}$,~\eqlabel{eq:binary_hubo} takes the form of a $p$-spin glass Hamiltonian~\cite{gardner1985spin},
\begin{equation}\label{eq:ising_hubo}
H_p = \sum_{i=1}^N h_i\,\sigma_i^z
+ \sum_{(m,n)\in G_2} J_{mn}\,\sigma_m^z\sigma_n^z
+ \sum_{(p,q,r)\in G_3} K_{pqr}\,\sigma_p^z\sigma_q^z\sigma_r^z,
\end{equation}
whose ground state is the optimal solution of $F(x)$. In the following sections, we describe procedures to generate these graphs and coefficients to construct classically hard instances that embed efficiently in quantum hardware.

\subsection{Instance generation}\label{sec:instances}

\begin{figure}[!t]
    \centering
    \includegraphics[width=\linewidth]{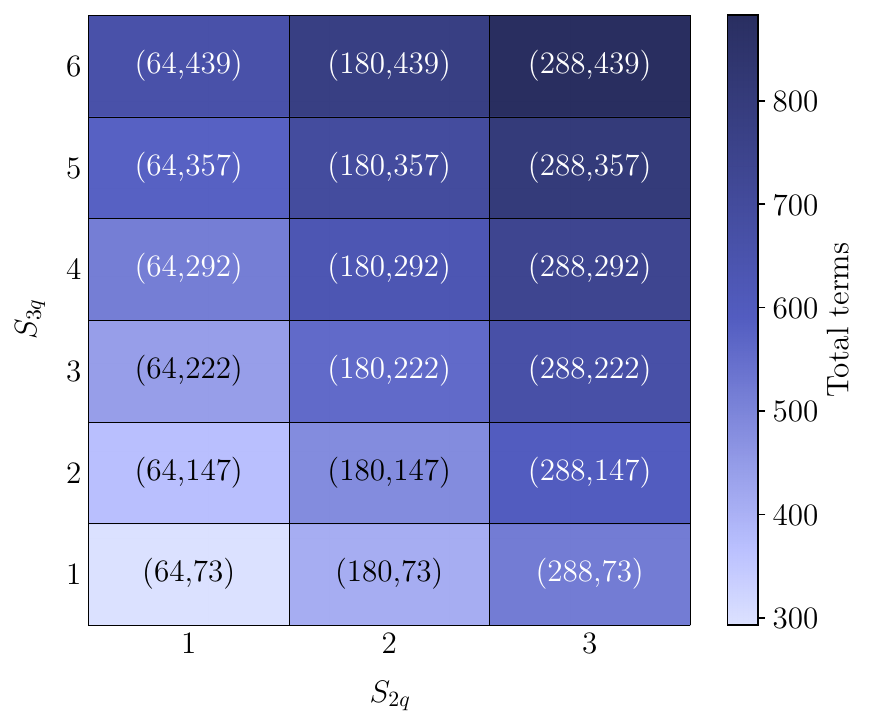}
    \caption{Number of two- and three-qubit interactions arranged as a tuple, for all the possible values of $S_{2q}$ and $S_{3q}$ with $n=1$ on a 156-qubit heavy-hexagonal architecture.}
    \label{fig:n_gates_for_1swap}
\end{figure}

We generate classically hard instances by jointly constructing the interaction graphs and selecting coupling coefficients. We start from empty graphs \(G_2 = \emptyset\) and \(G_3 = \emptyset\) and an initial coupling map \(C_0\) that encodes the hardware connectivity. Using graph‐coloring~\cite{kim2023evidence}, we identify all sets of nearest‐neighbor two‐ and three‐body terms that can be applied in parallel on \(C_0\). We denote these collections as $P_{2q} = \left\{P^{(1)}_{2q}, \cdots, P^{(M_2)}_{2q}\right\}$ and $P_{3q} = \left\{P^{(1)}_{3q}, \cdots, P^{(M_3)}_{3q}\right\}$, where $M_{2}$ $(M_{3})$ is the maximum number of two- (three-) body sets. Additionally, we define the integers $S_{2q}$, $(S_{3q})$ as the number of sets from $P_{2q}$ $(P_{3q})$ to include in $G_{2q}$ $(G_{3q})$, i.e. $G_{2q} \leftarrow G_{2q} \cup \left\{P^{(l)}_{2q}\right\}_{l=1,\cdots,S_{2q}}$. Similarly, $G_{3q} \leftarrow G_{3q} \cup \left\{P^{(l)}_{3q}\right\}_{l=1,\cdots,S_{3q}}$. 
Next, we use the first two‐body set $P_2^{(1)}$ as a \textsc{swap} layer, which permutes the qubit positions in $C_0$ and yields an updated coupling map $C_1$. The process is iteratively repeated for $n$ \textsc{swap} layers applied to the first set of $P_{2q}$,  see Algorithm~\ref{alg:swap_layers}. For all experiments, we choose $C_0$ to be the 156-qubit heavy-hexagonal lattice of IBM’s Heron architecture~\cite{chamberland2020topological}. Figure~\ref{fig:n_gates_for_1swap} shows the total number of Hamiltonian terms (one-, two-, and three-body interactions) as a function of $S_{2q}$ and $S_{3q}$ for $n=1$. As expected, increasing $S_{2q}$ and $S_{3q}$ raises the total interaction count from roughly $300$ to $800$.

\begin{figure}[htb]
\vspace{-2.7mm}
\begin{minipage}[t]{\columnwidth}%
\begin{algorithm}[H]%
\caption{Instance layout generation strategy}
\label{alg:swap_layers}
\begin{algorithmic}[1]
\Require Number of \textsc{Swap} layers $n$; initial hardware coupling map $C_0$; function \textsc{GraphColoring}$(C)$ that returns sets $P_{2q}$ and $P_{3q}$ of two- and three-qubit interactions executable in parallel respectively, given a coupling map $C$; function \textsc{SwapRegister}$(C,I)$ that swaps a set of pairs $I$ in the coupling map $C$, and returns the swapped coupling map.
\State Initialize $G_{2q} \gets \{\emptyset\}$ and $G_{3q} \gets \{\emptyset\}$
\State Set $C \gets C_0$
\For{$i = 1$ to $n$}
    \State $(P_{2q}, P_{3q}) \gets \textsc{GraphColoring}(C)$
    \State Select $S_{2q}$ subsets from $P_{2q}$ and append them to $G_{2q}$
    \State Select $S_{3q}$ subsets from $P_{3q}$ and append them to $G_{3q}$
    \If{$i<n$}
    \State $C \gets$ \textsc{SwapRegister}$\left(C,P^{(1)}_{2q}\right)$
    \EndIf
\EndFor
\State \Return $G_{2q}, G_{3q}$
\end{algorithmic}
\end{algorithm}
\end{minipage}
\end{figure}

After generating the interaction sets, we sample coupling coefficients from two heavy-tailed distributions: the Cauchy distribution and a variant of the Pareto distribution. Heavy-tailed distributions frequently produce large-magnitude coefficients, increasing the energy landscape’s ruggedness and making the combinatorial optimization problem harder to solve. Such heavy-tailed behavior is observed in real-world scenarios involving financial modeling~\cite{verhoeven2004fat, bollerslev2015tail, basnarkov2019option}.

\emph{Cauchy distribution.}---We draw the couplings from the standard Cauchy distribution, namely $T_{i_1\cdots i_k} \sim \text{Cauchy}(0,1)$. This distribution is characterized by the density function 
\begin{equation}
    f(\xi) = \frac{1}{\pi\left(1 + \xi^2\right)}.
\end{equation}
Since its distribution has no finite moments at any order, it is unsuitable for moment-based analysis. Nevertheless, its heavy-tailed nature often produces outlier-prone behavior, making it useful in robust statistics, signal processing, and noise modeling~\cite{ding2019total,karakus2020modelling}.

\emph{Pareto distribution.}---We draw the couplings from a symmetrized Pareto distribution, $T_{i_1\cdots i_k}\sim\mathrm{SymmetricPareto}(1,\alpha)$. This distribution is characterized by the density function 
\begin{equation}\label{eq:pareto}
f(\xi) = \left( \frac{\alpha}{\xi^{\alpha + 1}} + 1 \right) (-1)^y,
\end{equation}
where $\xi \geq 1$, and $y \in \{0, 1\}$ is a Bernoulli variable with 50\% probability. The term $\alpha /\xi^{\alpha + 1}$ corresponds to the standard Pareto distribution. This construction results in a heavy-tailed distribution centered at zero. For $\alpha = 2$, the mean exists and is finite, while the variance is undefined. This distribution produces occasional large-magnitude coupling coefficients, increasing the ruggedness and complexity of the optimization landscape. 
\begin{figure}[!tb]
    \centering
    \includegraphics[width=\linewidth]{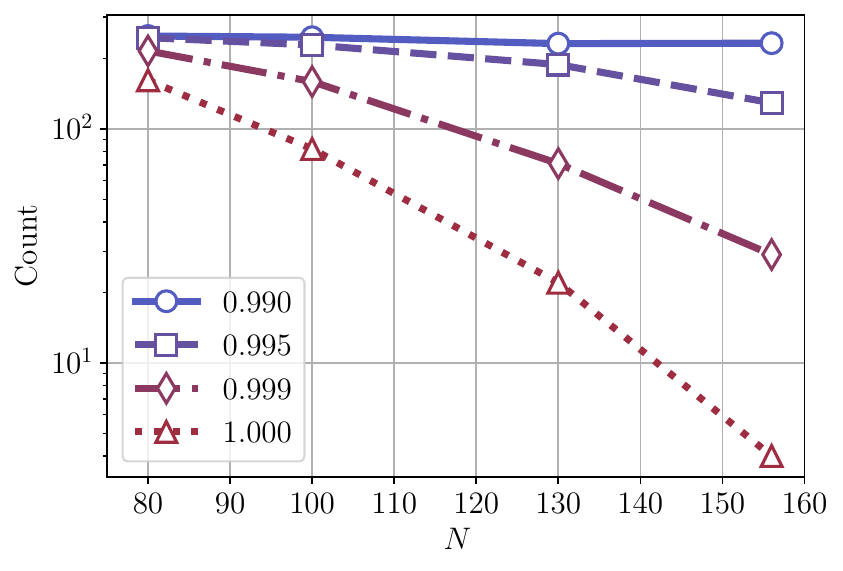}
    \caption{Number of instances where SA with $n_{\rm sweep}=20000$ and $n_{\rm runs}=100$ yield a solution within a given $\mathcal{R}$ for different system sizes $N$ between $80$ and $156$. The instances follow the Cauchy distribution with $n=1$, $S_{2q} = 1$ and $S_{3q}= 4$.}
    \label{fig:SA_250_cauchy_percentcloseness_N}
\end{figure}
\section{SA, CPLEX, and BF-DCQO}\label{sec:solvers}
To benchmark the performance of BF-DCQO, we consider two solvers: a heuristic classical solver based on the Metropolis-Hastings criterion~\cite{metropolis1953equation,hastings1970monte}, SA~\cite{kirkpatrick1983optimization}; an exact commercial solver, IBM CPLEX~\cite{cplex}. In this section, we provide a detailed description of the usage and implementation of each solver in detail.
\subsection{Simulated Annealing (SA)}
SA is a heuristic optimization technique inspired by the physical annealing process. The algorithm iteratively explores the solution space, accepting both energy-lowering and, with a certain probability, energy-increasing moves to escape local minima. As the algorithm progresses, the acceptance of higher energy moves decreases, guiding convergence toward near-optimal solutions.

In this work, we apply SA directly from the HUBO formulation of Eq.~\ref{eq:ising_hubo}. We initialize all spins randomly and independently compute the upper bound of the maximum energy change from any spin-flip $i$, namely $\Delta E^\text{max} = \max_i \Delta E_i^{\max}$. This sets the initial temperature, $T_{\text{init}} = \Delta E^\text{max}$, and we choose the final temperature as $T_{\text{final}} = 0.01\, T_{\text{init}}$. A geometric cooling schedule is generated between these bounds, assigning one temperature per sweep. Each SA run consists of $n_{\rm sweep}$ sweeps. In each sweep, we randomly permute the spin indices and visit them in that order. For each spin, we compute the exact energy change $\Delta E$ upon flipping and apply the Metropolis-Hastings criterion. After $n_{\rm sweep}$ sweeps, we record the lowest-energy configuration. We perform $n_{\rm runs}$ independent runs and return the best result. For all cases, SA is parallelized across all available CPU  cores, see Table.~\ref{tab:specs}, to maximize computational efficiency (see~\tablabel{tab:specs}). Defining $E$ as the lowest-energy valued sample and $E_\text{GS}$ the most optimal solution obtained using CPLEX, Figure~\ref{fig:SA_250_cauchy_percentcloseness_N} shows how many instances out of 250 achieve an approximation ratio $\mathcal{R}=E/{E_{\rm GS}}$ between $0.99$ and $1.00$ as a function of system size $N$ ranging from $80$ to $156$, and using $n_{\rm sweep}=20000$ and $n_{\rm runs}=100$. For reference, the total runtime of SA is computed using the expression
$T_{\rm{CPU}} = n_{\rm{sweep}}\, n_{\rm{runs}} \cdot 0.6\cdot 10^{-5} s$, where $ 0.6\cdot 10^{-5} s$ denotes the time per sweep, obtained by extrapolating the average runtimes of zero-temperature SA across a range of $10^4$ to $10^7$ sweeps, see Appendix~\ref{App:runtimes}.We find that smaller $S_{2q}$ increases problem difficulty (see Appendix~\ref{App:Cauchy}).

Based on this observation, we set $S_{2q}=1$ and $S_{3q}=4$ for our subsequent analysis. This choice generates compact circuits while preserving problem difficulty for experimental implementations. We sample the coupling coefficients from a Cauchy distribution with values constrained to the range $[-7, 7]$. As $N$ grows, fewer instances reach high approximation ratios $\mathcal{R}$, indicating increased problem hardness and making it more difficult for SA to reach close-to-optimal solutions (see Appendix~\ref{App:Cauchy}). Here, we also observed that CPLEX (detailed in the next section) can solve these instances within a few seconds on average. Therefore, we use CPLEX results as the baseline for comparing SA and BF-DCQO. 
\begin{figure}[!tb]
    \centering
    \includegraphics[width=1\linewidth]{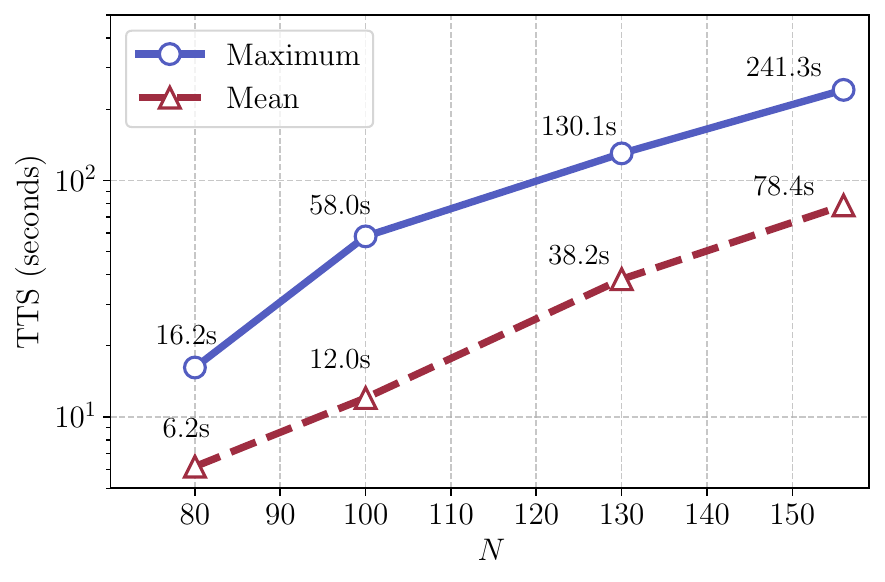}
    \caption{Maximum and mean TTS of CPLEX for $250$ randomly generated instances as a function of the system size $N$. The instances follow the symmetrized Pareto distribution with $\alpha=2$, $n=1$, $S_{2q}=1$, and $S_{3q}=6$.}
    \label{fig:tts_cplex_N}
\end{figure}
\subsection{CPLEX}
CPLEX is a commercial solver owned by IBM that targets mixed-integer linear and quadratic programming. It is widely used in large-scale optimization and serves as a robust classical benchmark. All results were obtained using the software specifications in Table~\ref{tab:specs}. To solve our HUBO problems using CPLEX, we convert each instance to a mixed-integer programming problem. Several conversion methods are described in Appendix~\ref{App:CPLEXconverter}; we selected the method with the lowest runtime on our generated instances.
\begin{table}[!tb]
    \caption{Classical hardware and software specifications.}\label{tab:specs}
    \begin{ruledtabular}\begin{tabular}{rlrl}
       Processor & \multicolumn{3}{l}{AMD (KVM processor)  ($48\times\SI{2.3}{GHz}$)} \\
       RAM & \multicolumn{3}{l}{$\SI{123}{GB}$} \\
       OS & \multicolumn{3}{l}{Debian GNU/Linux 12 (bookworm) ($\times 64$)} \\
       CPLEX~\cite{cplex} & v22.1.2.0 & C++ & 11.4.0 \\
    \end{tabular}\end{ruledtabular}
\end{table}
\begin{figure*}[!tb]
    \centering
    \includegraphics[width=\linewidth]{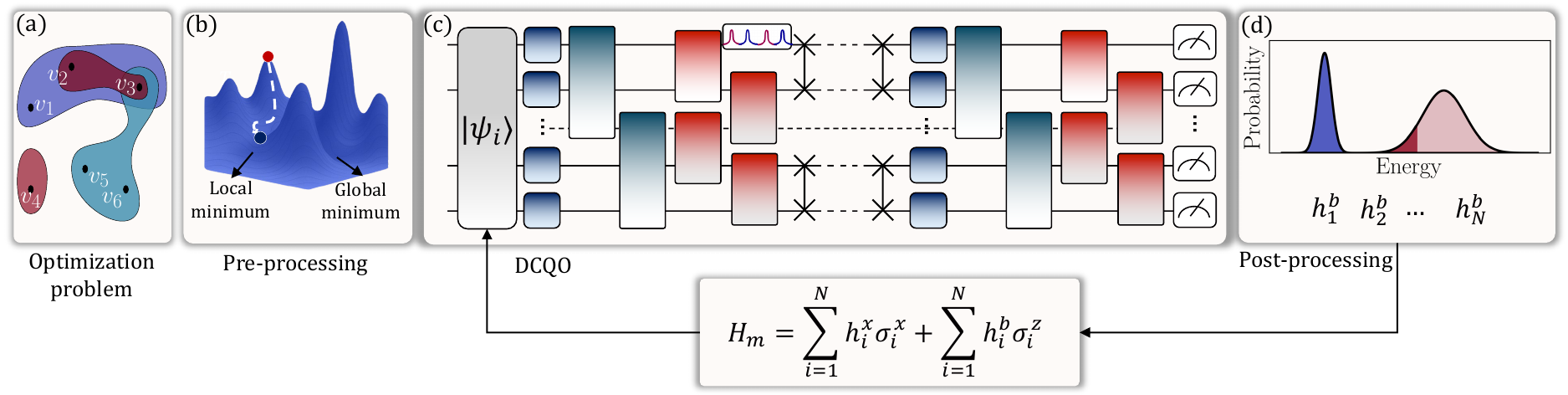}
    \caption{Workflow of the BF-DCQO algorithm. (a) A HUBO problem is generated as a hypergraph following the method described in Sec.~\ref{sec:instances}. Then, they are mapped to an Ising Hamiltonian $H_p$ whose ground state encodes the solution. (b) Pre-processing: SA with a fixed number of sweeps $n^{\rm{pre}}_{\rm{sweep}}$ is then applied to reach a regime where classical approximation becomes difficult. (c) BF-DCQO: The lowest-energy bitstrings from SA are used to initialize the bias fields. Then, the DCQO circuit is constructed using the first-order approximate adiabatic gauge potential $A_{\lambda}^{(1)}$, implemented as a single Trotter step ($n_{\rm{trot}}=1$). Single-qubit gates are executed in parallel, followed by parallelized three- and two-body terms, and a \textsc{swap} layer. This structure is repeated for $n$ \textsc{swap} layers. A dynamical decoupling sequence (shown by red and blue pulses) is also applied to mitigate errors from idling qubits. The circuit is then transpiled into hardware-native gates. Finally, measurements are taken in the computational basis. (d) Post-processing: From the sampled bitstrings, the $n_{\rm{CVaR}}$ lowest-energy states are selected (red), and zero-temperature simulated annealing with $n^{\rm{post}}_{\rm{sweep}}$ sweeps is applied, producing an improved distribution (blue). Bias fields are then computed from this refined distribution, a new mixer Hamiltonian $H_m$ is constructed, and its ground state is prepared for the next iteration. This process is repeated for $n_{\rm{iter}}$ iterations until a high-quality approximate solution is obtained.}
    \label{fig:schematic}
\end{figure*}

In Figure~\ref{fig:tts_cplex_N}, we measure the time to reach a provably optimal solution, denominated as TTS, for CPLEX over $250$ random Pareto-distributed HUBO instances. We restricted to system sizes $N$ between $80$ and $156$ and fixed $n=1$, $S_{2q}=1$, and $S_{3q}=6$, which maximizes TTS, as described in Appendix~\ref{App:Pareto}. For all the cases, we used a single CPU thread and observed an exponential scaling of TTS with respect to $N$. This exponential growth highlights the practical limits of classical solvers and motivates alternative approaches. We also observe that for these instances, SA can find the optimal solution with a few thousand sweeps.
\begin{figure*}
    \centering
    \includegraphics[width=1\linewidth]{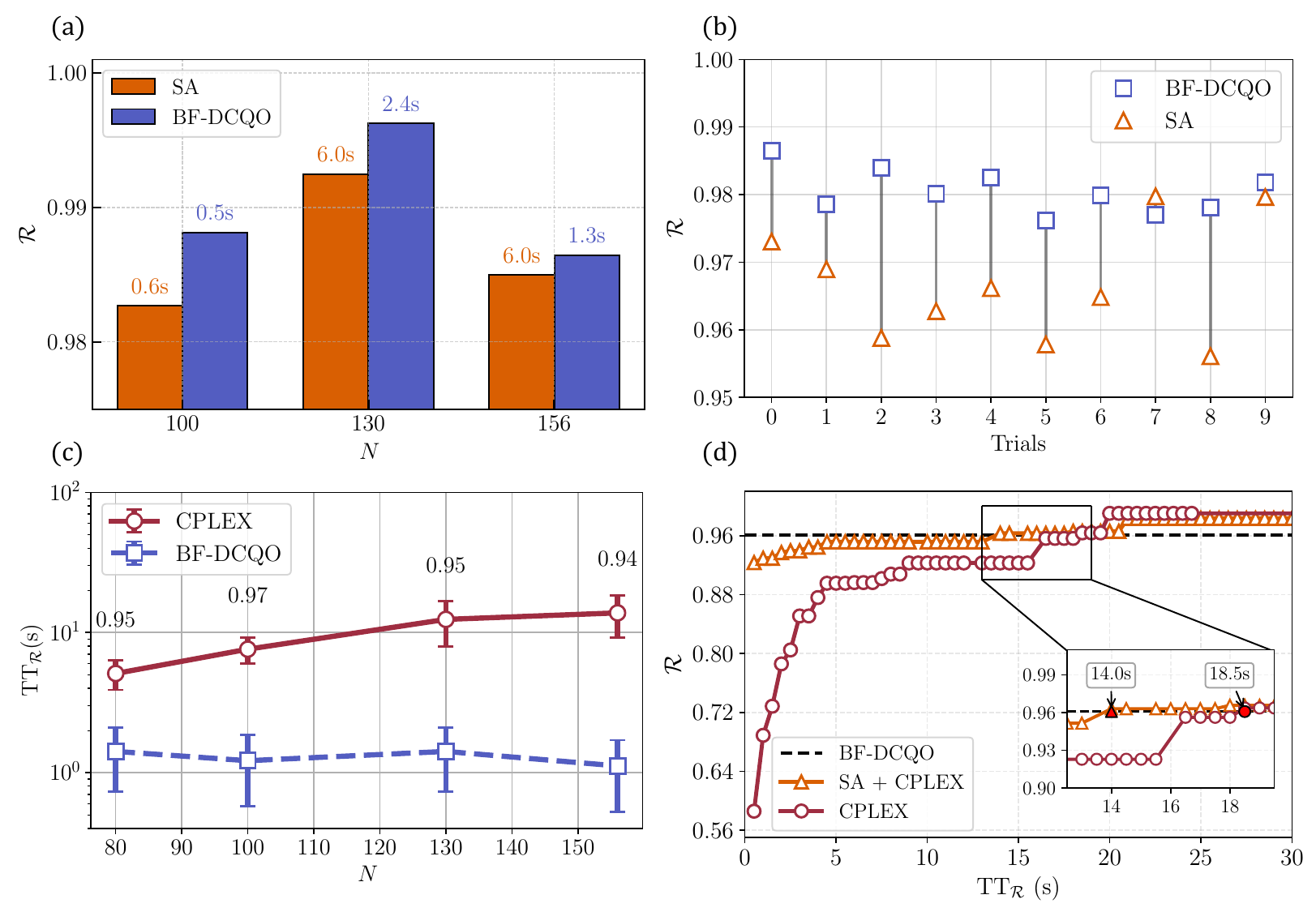}
    \caption{Experimental results. 
    (a) Approximation ratio $\mathcal{R}$ for SA and BF-DCQO as a function of the number of qubits $N$, ranging from $N=100$ to $N=156$. The bars represent the $\mathcal{R}$ values of the best instance in terms of the performance of BF-DCQO, across 5 problem instances for each $N$ for SA and BF-DCQO, respectively, with the $\mathrm{TT}_\mathcal{R}$ values mentioned respectively. (b) Approximation ratio $\mathcal{R}$ as a function of trials corresponding to 10 different experiments of BF-DCQO with around 1.3 seconds runtime, and 10 runs of SA with 100,000 sweeps with 6 seconds of runtime per run. The experiments were performed on the $N=156$ qubit instance corresponding to the Cauchy-distributed class. (c) TT$_\mathcal{R}$ for CPLEX and BF-DCQO as a function of the number of qubits $N$, ranging from $N=80$ to $N=156$. Solid lines represent the mean TT$_\mathcal{R}$ across 5 problem instances for each $N$. TT$_\mathcal{R}$ for CPLEX denotes the time taken to reach the minimum energy obtained by BF-DCQO. 
    (d) Approximation ratio $\mathcal{R}$ as a function of time for the best-performing instance with $N=156$ qubits, comparing CPLEX and BF-DCQO. The red line shows the $\mathcal{R}$ obtained by CPLEX over time, with the BF-DCQO energy level indicated by a dashed line. The red line shows the CPLEX run with the SA-derived solution as initialization. The intersection point highlights the time at which CPLEX first reaches the energy achieved by BF-DCQO. The inset zooms in on the convergence region between 12 and 20 seconds, illustrating when CPLEX matches the BF-DCQO energy level.}
    \label{fig:Results_expt}
\end{figure*}
\subsection{BF-DCQO}
The digitized counterdiabatic quantum optimization (DCQO) algorithm implements a digitized fast evolution governed by a CD Hamiltonian $H_{cd}$, which is built from a finite-time adiabatic path from a mixer Hamiltonian $H_m$ to the target Hamiltonian $H_p$. In BF-DCQO~\cite{cadavid2024bias,romero2024bias,iskay}, the DCQO algorithm (see Appendix~\ref{App:dcqc}) is applied iteratively by biasing $H_m$ as 
\begin{equation}
    H_m = \sum_{i=1}^N h^x_i \sigma^x_i +  \sum_{i=1}^N h_i^b \sigma^z_i,
\end{equation}
where the transverse fields $h^x_i$ are kept constant, and the longitudinal ``bias" fields $h^b_i$ are updated iteratively using the update rule $h^b_i = \pm\langle \sigma^z_i \rangle$. This measurement-based feed-forward approach enables the preparation of progressively improved initial states at each iteration. Furthermore, we employ the first-order nested commutator expansion of the adiabatic gauge potential (see Appendix~\ref{App:dcqc}), where $H_{cd}$ takes the form
\begin{equation}\label{eq:o1}
\begin{split}
H_{cd} &= -2\beta_1(t) \Bigg[
     \sum_{i=1}^N h_i^{x} h_i^z \,\sigma_i^y 
    + \sum_{(m,n) \in G_{2q}} J_{mn} \left( h_m^{x} \sigma_m^y \sigma_n^z + h_n^{x} \sigma_m^z \sigma_n^y \right) \\
    & + \sum_{(p,q,r) \in G_{3q}} K_{pqr} \Big( h_p^{x} \sigma_p^y \sigma_q^z \sigma_r^z 
    + h_q^{x} \sigma_p^z \sigma_q^y \sigma_r^z + h_r^{x} \sigma_p^z \sigma_q^z \sigma_r^y \Big)
\Bigg],
\end{split}
\end{equation}
with  the analytical form of $\beta_1(t)$ provided in Ref.~\cite{romero2024bias}. At each iteration, the ground state of $H_m$ is prepared as $\ket{\psi_i} = \bigotimes_{i=1}^N R_y(\theta_i) \ket{0}$, where $\theta_i = \tan^{-1} \left( \frac{h^x_i}{h_i^b + \sqrt{(h_i^b)^2 + (h_i^x)^2}} \right)$. Then, the system evolves in the impulse regime under the time-evolution operator generated by $H_{cd}$, which includes one-, two-, and three-body terms. This evolution circuit is implemented in the following order, it begins with all single-qubit gates applied in parallel, followed by parallel layers of three-body and two-body interaction terms, and then a \textsc{swap} layer. This sequence is repeated for $n$ \textsc{swap} layers, with the application of only $S_{3q}$ three-body and $S_{2q}$ two-body terms, c.f. Algorithm~\ref{alg:swap_layers}. This ordering ensures a compact compilation, and the placement of three-body terms before two-body terms is deliberate, as subsequent \textsc{swap} operations can further simplify the circuit by canceling redundant gates. Next, we measure $n_{\rm{shots}}$ times in the computational basis. From the resulting distribution, the $n_{\rm{CVaR}}$ lowest-energy states are selected, analogous to the conditional-value-at-risk (CVaR) strategy~\cite{Barkoutsos2020improving,barron2023provableboundsnoisefreeexpectation,romero2024bias}. The corresponding $h^b_i$ values associated with this reduced state are then used to update $H_m$ and $H_{cd}$. This iterative procedure is repeated for $n_{\rm{iter}}$ iterations. 

Additionally, we incorporate classical pre- and post-processing steps to further enhance both the bias initialization and the obtained solution quality. The pre-processing step consists of executing SA with a low number of sweeps and runs. Then, the lowest energy bitstring is used to initialize the bias fields. On the other hand, we employ zero-temperature SA with $n^{\rm{post}}_{\rm{sweep}}$ on the reduced states as post-processing, where each sweep involves flipping individual bits and accepting the change only if it leads to a lower energy. This post-processing step serves to recover information potentially lost due to random bitflip errors during experimentation, thereby enhancing the overall solution quality and robustness. The entire BF-DCQO algorithm is summarized in Figure~\ref{fig:schematic}.

The total runtime for BF-DCQO, including both classical and quantum resources, is computed as $T_{\rm BF-DCQO} = T_{\rm{CPU}} + T_{\rm{QPU}}$, where $T_{\rm{CPU}} = \left[n_{\rm{sweep}}^{\rm pre}  n_{\rm{runs}} 
+ n_{\rm{CVaR}} (n_{\rm{iter}}{+}1) n_{\rm{sweep}}^{\rm post} \right] 0.6 \cdot \SI{e-5}{s}$ and $T_{\rm{QPU}} = (n_{\rm{iter}}{+}1) n_{\rm{shots}} \cdot \SI{e-4}{s}$, respectively. Here the term $n^{\rm{pre}}_{\rm{sweep}}\, n_{\rm{runs}}$ accounts for the classical pre-processing, while $n_{\rm{CVaR}}\, (n_{\rm{iter}}{+}1)\, n^{\rm{post}}_{\rm{sweep}}$ for the post-processing. For the quantum runtime, we assume a shot rate of $10^4$ shots per second, which corresponds to the repetition rate configured on the quantum hardware. Note that circuit compilation time is excluded, since all the instances are built on heavy-hexagonal lattice with fixed $S_{2q}$ and $S_{3q}$ values. Hence, the circuits can be parameterized and pre-compiled.

\section{Experimental Results}\label{sec:results}
In this section, we present and analyze the performance of BF-DCQO in comparison to SA and CPLEX, demonstrating how BF-DCQO can achieve better approximate solutions experimentally for certain problem instances. We employ time-to-approximate solution (TT$_\mathcal{R}$) as a performance metric, defined as the time required to reach a solution that has approximation ratio $\mathcal{R}$ to the optimum. This time is measured in seconds and takes into account the algorithmic resources as well as their physical times. For BF-DCQO, we fix the algorithmic parameters to make a consistent and systematic performance benchmark. 

The experiments were performed on the 156-qubit \textsc{IBM Marrakesh} quantum processor, accessed via cloud using Qiskit~\cite{Javadi-Abhari2024Quantum}. The BF-DCQO circuits were transpiled using the Qiskit transpiler with optimization level 3 into the hardware-native gate set $\{\textsc{CZ}, R_z(\theta), \sqrt{X}, X\}$, where $\text{CZ}=\text{diag}(1,1,1,-1)$ and $R_z(\theta)=\exp(-i\theta\sigma^z/2)$. Additionally, we utilized fractional gates~\cite{ibm_fractional_gates} that are natively supported on IBM's Heron QPUs whenever possible. Specifically, $R_{zz}(\theta)=\exp(-i\theta\sigma^z_0\sigma^z_1/2)$ for $0 < \theta \leq \pi/2$ and $R_x(\theta)=\exp(-i\theta\sigma^x/2)$ gates for arbitrary $\theta$, enabling more efficient implementation of entangling operations at a reduced circuit depth. Furthermore, we enabled dynamical decoupling with the \texttt{XpXm} pulse sequence to suppress decoherence effects, setting the scheduling method to as-soon-as-possible (asap) and a slack distribution strategy centered in the middle of idle windows.

\subsection{Outperforming SA}
We compare BF-DCQO and SA on up to five hard Cauchy-distributed HUBO instances for system sizes $N=100,130,156$ taken from Figure~\ref{fig:SA_250_cauchy_percentcloseness_N}. For the $N=100$ qubit instances, we initialize BF-DCQO using the best bitstring from SA performed with $n_{\rm sweep}=500$ and $n_{\rm runs} = 100$. The motivation here is to quickly approach the vicinity of the optimal solution using SA and then exploit BF-DCQO to tunnel through local minima to obtain improved approximate solutions. Since SA is known to perform well at smaller system sizes, we fix $n_{\rm{iter}} = 0$ for these instances, effectively using SA-initialized DCQO with no further bias-field updates. For classical post-processing, we set $n_{\rm{CVaR}} = 100$ and $n^{\rm{post}}_{\rm{sweep}} = 10$.

Figure~\ref{fig:Results_expt}c shows the approximation ratio $\mathcal{R}$ for SA and BF-DCQO on the best-performing instance on hardware and the time ${\rm TT}_\mathcal{R}$ taken. At $N=100$, BF-DCQO achieves a lower $\mathrm{TT}_\mathcal{R}$ than SA ($n_{\rm sweeps} =1000$, $n_{\rm runs}=100$) in $\SI{0.5}{s}$, whereas SA took $\SI{0.6}{s}$. For the $N=130$ and $N=156$ qubit instances, the selected problems are significantly more challenging for SA. Even with 6 seconds of runtime ($n_{\rm sweep} = 10000$ and $n_{\rm reads} = 100$), SA fails to reach equal solution quality as BF-DCQO. For these two system sizes, we allow up to $n_{\rm{iter}}=3$, with $n^{\rm pre}_{\rm{sweep}}=1000$ and $n_{\rm runs} = 100$. This shows BF-DCQO is able to reach high-quality approximate solutions more efficiently, escaping local minima that hinder classical heuristics. Check the experimental details in Appendix~\ref{App:additional_SA}.

We assess robustness by selecting the best-performing $N=156$ instance with BF-DCQO in terms of $\mathcal{R}$ ten times on the quantum hardware, each time using the same SA-initialized bitstring. Figure~\ref{fig:Results_expt}d compares the minimum energies from these BF-DCQO runs to those from SA with $100000$ sweeps ($\SI{6}{s}$ runtime). Remarkably, BF-DCQO achieved a lower energy in $9$ of $10$ trials while using less than half the time of SA. This demonstrates that BF-DCQO is reliable across repeated experiments.

\subsection{Outperforming CPLEX}
We benchmark BF-DCQO against CPLEX on five Pareto-distributed HUBO instances for $N=80,100,130,156$, taken from Figure~\ref{fig:tts_cplex_N}. BF-DCQO parameters are set to $n_{\rm trot}=1$, $n_{\rm CVaR}=100$, $n^{\rm pre}_{\rm reads}=1$ and $n^{\rm post}_{\rm sweep}=10$. For $N=80$, we set $n^{\rm{pre}}_{\rm{sweep}}=10$ and for $N\geq 100$ we use $n^{\rm{pre}}_{\rm{sweep}}=100$. In Figure~\ref{fig:Results_expt}a, we show the distribution of $\mathrm{TT}_\mathcal{R}$. We observe that BF-DCQO reaches $\mathcal{R}\approx0.95$ in just a few seconds, demonstrating its ability to deliver high-quality approximate solutions. To compare with CPLEX, we first run BF-DCQO to achieve a certain $\mathcal{R}$ values, then we run CPLEX to check how much time it takes to  Besides, the gap between BF-DCQO and CPLEX widens with increasing system size. This gap arises because CPLEX must reformulate HUBO as mixed-integer programs, adding variables and constraints, which makes its inherent relaxations ineffective and requires deep exploration in the search tree to start converging, consequently slowing down its search. All results use instances with a single-\textsc{swap} layer. Adding more swap layers would greatly increase CPLEX’s runtime and push circuit depth for BF-DCQO beyond current coherence-time limits, so we focus on this single-swap-layer regime. We expect that hardware improvements, such as longer coherence times, lower gate errors, or more qubits will further boost BF-DCQO’s performance on HUBO problems.

To ensure a fair comparison and to verify that the performance advantage arises from the quantum component of the entire workflow, we ran CPLEX on all considered instances using the same SA-derived bitstring as a warm start, identical to the initialization used in BF-DCQO. The corresponding results are detailed in Appendix~\ref{App:additional_SA}. In Figure~\ref{fig:Results_expt}b, we illustrate the evolution of energy over time for both CPLEX and SA-initialized CPLEX for the best-performing instance corresponding to $N=156$ qubits, where BF-DCQO reaches its solution in approximately $0.2$ seconds. We observe that the energy obtained by CPLEX initially decreases rapidly but eventually enters a plateau region. This behavior is likely due to the internal sub-algorithms of CPLEX becoming trapped while exploring a subset of solutions, resulting in a stagnation of improvement. Despite this, it takes approximately 17 seconds for CPLEX to reach the energy level achieved by BF-DCQO. Similarly, SA-initialized CPLEX begins from a lower energy state, but requires around $14$ seconds to reach the corresponding BF-DCQO solution. This relative inefficiency highlights the strength of BF-DCQO, which can traverse the solution landscape much more efficiently within a shorter runtime window. 
\begin{figure}[t]
    \centering
    \includegraphics[width=1\linewidth]{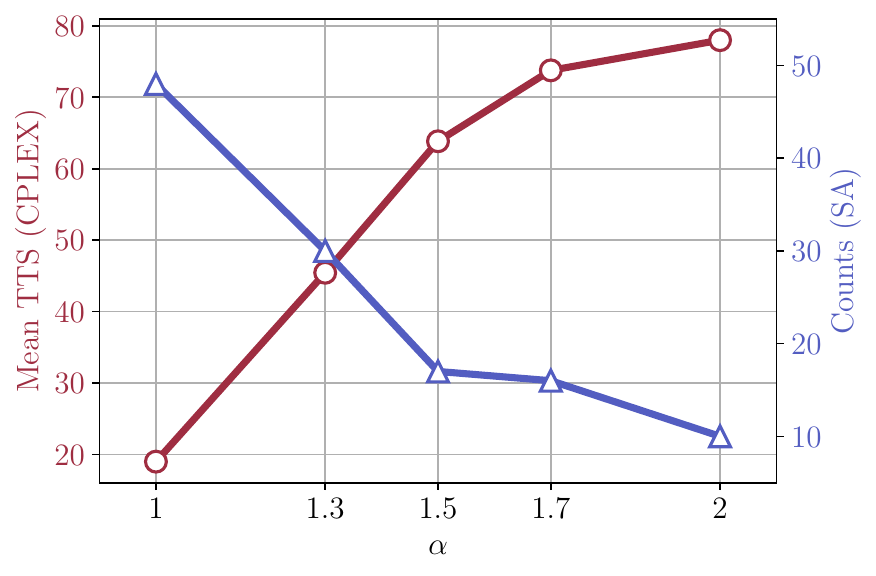}
    \caption{Comparison of CPLEX and SA performance as a function of Pareto shape parameter $\alpha$ for $N=156$ qubits with $S_{2q}=1$, $S_{3q}=6$, and $n=1$ \textsc{swap} layer. The left axis shows the mean TTS (in seconds) required by CPLEX to solve 50 randomly generated instances. The right axis indicates the number of instances for which SA failed to find the optimal solution after $n_{\rm{sweep}} = 20000$ and $n_\text{runs}=100$.}
    \label{fig:tts_vs_sa_quality}
\end{figure}%
\begin{figure*}[t]
    \centering
    \includegraphics[width=1\linewidth]{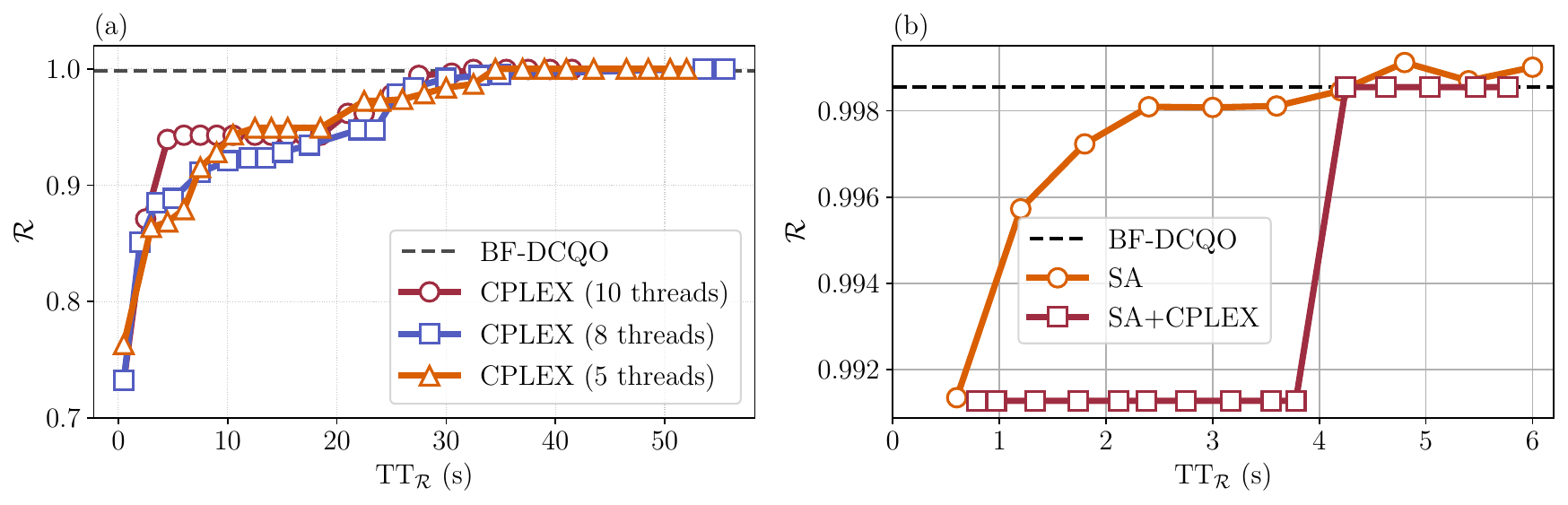}
    \caption{(a) Approximation ratio ($\mathcal{R}$) as a function of runtime for CPLEX with increasing number of threads from 5 to 10. (b) Approximation ratio ($\mathcal{R}$) as a function of runtime for SA and warm-started CPLEX, compared with the BF-DCQO solution. The black dashed line indicates the $\mathcal{R}$ value achieved by BF-DCQO in TT$_\mathcal{R} = 0.506$ seconds. The SA curve shows the mean $\mathcal{R}$ values obtained over $10$ trials, each consisting of $n_\text{runs}=100$ and $n_{\rm{sweep}} \in [1000, 10000]$. The SA+CPLEX curve represents CPLEX performance when initialized with the same SA-derived solution used for BF-DCQO.}
    \label{fig:beating_both_ar_vs_time}
\end{figure*}

\subsection{Outperforming both SA and CPLEX}
Until this point, we analyzed instances that are hard to solve for either SA or CPLEX.  Out of $250$ instances, we selected some of the hardest ones and executed BF-DCQO on IBM hardware. We also observed that the instances that are hard for CPLEX are easy for SA and vice versa. Here, we extend the analysis to instances where there is potential to outperform both solvers simultaneously. In the previous sections, the shape parameter for the Pareto distributed instances was fixed at $\alpha = 2$ for generating the coefficients of the problem Hamiltonian $H_p$. Here, we treat $\alpha$ as a tunable parameter and vary it from $\alpha = 1$ to $\alpha = 2$. As the shape parameter $\alpha$ increases in the Pareto distribution, the tail of the distribution becomes thinner, reducing the likelihood of extreme values. For $\alpha \leq 1$, the mean is undefined. For $1 < \alpha \leq 2$, the mean exists but the variance remains infinite. When $\alpha > 2$, both the mean and variance are finite, and the distribution concentrates around lower values. Therefore, increasing $\alpha$ yields a distribution that is less heavy-tailed and more statistically stable. 

In Figure~\ref{fig:tts_vs_sa_quality}, we compare the performance of CPLEX and SA by generating $50$ random instances with varying values of $\alpha$ for a system size of $N=156$ qubits, with $S_{2q}=1$, $S_{3q}=6$, and $n=1$ \textsc{swap} layer. For CPLEX (left axis), we plot the mean TTS in seconds required to solve the problem. For SA (right axis), we show the number of instances in which SA failed to find the optimal solution after $n_{\rm{sweep}}=20000$ and $n_{\rm runs}=100$. We observe that as the shape parameter $\alpha$ increases, the mean TTS values for CPLEX increase from approximately $20$ seconds for $\alpha = 1$ to about $80$ seconds for $\alpha = 2$. This indicates that, for our class of instances, extremely heavy-tailed distributions are not necessary to construct problems that are challenging for CPLEX to solve as compared to standard distributions such as Gaussian, where CPLEX tends to solve the problems much faster. In contrast, the behavior for SA shows an opposite trend. As $\alpha$ increases, the number of instances for which SA reaches the optimal cost at least once decreases significantly, from around $50$ instances at $\alpha = 1$ to only $10$ at $\alpha = 2$. This suggests that lower $\alpha$ values, corresponding to heavier-tailed distributions, generate instances that are more difficult for SA to solve.

Therefore, this complementary behavior implies that while lower $\alpha$ values are suited to generating hard instances for SA, higher $\alpha$ values tend to yield harder instances for CPLEX. By carefully tuning $\alpha$, one can generate problem instances that are simultaneously difficult for both solvers, thus providing a robust testbed for benchmarking. It is also important to note that these trends represent mean values; individual instances may deviate from the average, and in some cases, the hardest instance for CPLEX may also pose a significant challenge for SA. This analysis enables better control over instance hardness and increases the likelihood of generating suitably difficult problems.

To demonstrate a runtime advantage over both CPLEX and SA simultaneously, we fixed the Pareto shape parameter to $\alpha=2$, as in previous experiments, but selected the hardest instance out of a pool of $250$ generated instances. Since this instance is challenging for both SA and CPLEX. We employed a stronger pre-processing step to move the solution into a region where SA begins to struggle. For BF-DCQO, we used $n^{\rm{pre}}_{\rm{sweep}}=500$, $n_\text{reads}^\text{pre} = 100$, $n_{\rm{iter}}=0$ and $n_{\rm{shots}}=2000$, while keeping all other hyperparameters unchanged from earlier experiments. The implementation was carried out on the \textsc{IBM Kingston} backend without the use of fractional gates.

In Figure~\ref{fig:beating_both_ar_vs_time}a, we compare runtimes of CPLEX using 5, 8, and 10 threads as it converges to the approximation ratio $\mathcal{R}$ achieved by BF-DCQO in TT$_\mathcal{R}=0.506\,$s. We observe that CPLEX requires approximately $\SI{34}{s}$ with 5 threads, $\SI{51}{s}$ with 8 threads, and $\SI{31}{s}$ with 10 threads to reach this performance. We note that increasing the number of threads further doesn't decrease the runtime further. To enable a fair comparison, we measured TT$_\mathcal{R}$ for CPLEX initialized with the same SA-derived bitstring, and for SA executed with sweep counts ranging from $n_{\rm{sweep}} \in [1000, 10000]$ and $n_\text{reads} = 100$. The results, summarized in Figure~\ref{fig:beating_both_ar_vs_time}b, show the mean $\mathcal{R}$ over $10$ independent trials for both CPLEX and SA. The black dashed line indicates the best energy obtained by BF-DCQO, which was reached in TT$_\mathcal{R} = \SI{0.506}{s}$. In contrast, SA alone required approximately $4$ seconds (corresponding to $n_{\rm{sweep}}=7000$) to reach the same solution quality, while CPLEX initialized with SA also needed around $4$ seconds on average. It is worth noting that, without the warm start, CPLEX would require significantly longer to converge to this solution. These results present experimental evidence that BF-DCQO can outperform both tested classical solvers in runtime simultaneously.

\section{Conclusion}\label{sec:conclusion}

In this work, we have given experimental evidence that a runtime quantum advantage from a quantum algorithm with respect to specific classical algorithms is possible for approximate solutions of HUBO problems. We have developed a protocol to generate problem instances that are hard for specific classical methods, yet feasible on current quantum devices. We have benchmarked the performance of bias-field digitized counterdiabatic quantum optimization (BF-DCQO) against both SA and CPLEX, finding cases of runtime advantage over each classical method under specific conditions. This was accomplished by generating up to three-body HUBOs through systematic application of \textsc{swap} layers and parallel interactions, resulting in densely connected instances compatible with the \textsc{IBM Heron} heavy-hexagonal architecture. The problem instances contain heavy-tailed distributed coefficients, which are challenging for classical solvers. Particularly, we used Cauchy-distributed coefficients to create difficult instances for SA, while Pareto-distributed coefficients for CPLEX. Additionally, we found that tuning the $\alpha$ parameter from the Pareto distribution could result in difficult instances for both solvers. Leveraging this insight, we generated several instances with system sizes ranging from $N=80$ to $N=156$ qubits, many of which required several minutes of CPU time to be solved to optimality classically. We then applied the BF-DCQO experimentally and demonstrated that it can reach approximate solutions significantly faster than tested classical approaches. For each system size ($N=80$, $100$, $130$, and $156$), we considered experimentally five instances and observed that BF-DCQO can reduce runtimes by up to a factor of $80$ compared to CPLEX. Complementary to that, we also observed more than a $3.5\times$ reduction in runtime for BF-DCQO compared to SA in the best-case scenario.

All experiments were conducted using only a single \textsc{swap} layer, suggesting that increasing the interaction depth would make the classical runtimes even longer, potentially reaching hours or days, which leaves BF-DCQO at a favorable position to remain comparatively efficient. This study illustrates the practical utility of current quantum hardware without the need for quantum error correction. It also provides experimental evidence for heuristic digital quantum optimization speedups, pointing toward the possibility of quantum advantage. Ultimately, our results show that contemporary quantum processors, when paired with advanced algorithms like BF-DCQO, could be capable of delivering solutions to industrial-scale optimization problems.

\begin{acknowledgments}
    We thank Stefan Woerner for his insightful comments and validation of our benchmark test with CPLEX as well as valuable feedback. We acknowledge the use of IBM Quantum services for this work. The views expressed are those of the authors and do not reflect the official policy or position of IBM or the IBM Quantum team.
\end{acknowledgments}
\appendix

\section{HUBO reduction techniques}\label{App:CPLEXconverter}
\subsection{HUBO to MIP}
To solve HUBO problems using classical optimization techniques, we convert them into equivalent mixed-integer programming (MIP) formulations~\cite{winston2004operations}. The HUBO Hamiltonian includes linear, quadratic, and higher-order terms defined over spin variables $s_i \in \{-1, +1\}$. These spin variables are first mapped to binary variables $x_i \in \{0, 1\}$ via the substitution $s_i = 1 - 2x_i$. The next step is to linearize all higher-order monomials to obtain a formulation suitable for MIP solvers.

\emph{Quadratic terms.}---Each quadratic term $x_i x_j$ is replaced by a binary auxiliary variable $a_{ij}$ that represents the product. To ensure the equivalence $a_{ij} = x_i x_j$, the following constraints are introduced:
\begin{align}
    a_{ij} &\le x_i, \\
    a_{ij} &\le x_j, \\
    a_{ij} &\ge x_i + x_j - 1.
\end{align}
These inequalities enforce that $a_{ij} = 1$ if and only if both $x_i$ and $x_j$ are 1.

\emph{Cubic terms.}---Cubic terms of the form $x_i x_j x_k$ are handled in two stages. First, an auxiliary variable $a_{ijk}$ is introduced to encode the product $x_i x_j$ using
\begin{align}
    a_{ijk} &\le x_i, \\
    a_{ijk} &\le x_j, \\
    a_{ijk} &\ge x_i + x_j - 1.
\end{align}
Next, a second auxiliary variable $b_{ijk}$ is introduced to represent the final cubic product via $b_{ijk} = a_{ijk} x_k$, enforced by
\begin{align}
    b_{ijk} &\le a_{ijk}, \\
    b_{ijk} &\le x_k, \\
    b_{ijk} &\ge a_{ijk} + x_k - 1.
\end{align}
These constraints ensure that $b_{ijk} = 1$ if and only if all three variables $x_i$, $x_j$, and $x_k$ are equal to 1.

\emph{Final MIP formulation.}---The final MIP formulation consists of binary variables $x_i$ corresponding to each original spin variable in the HUBO, one auxiliary variable, and three linear constraints for each quadratic term, and two auxiliary variables and six linear constraints for each cubic term. The objective function is expressed as a linear function involving the constant offset, linear contributions from the $x_i$ variables, and terms arising from the auxiliary variables corresponding to the quadratic and cubic interactions. This fully linearized binary MIP model can then be optimized using standard solvers such as CPLEX or Gurobi.

\emph{Warm-starting CPLEX.}---Since there is a pre-processing step for BF-DCQO, we also provide the MIP solver with the same initial solution, commonly known as a warm start. This approach can be particularly effective for MIP problems, where the search space is combinatorially large and early guidance can significantly influence solver efficiency. A warm start consists of supplying the solver with a candidate solution that satisfies all model constraints, enabling it to initiate the optimization from a known feasible point rather than relying entirely on internal heuristics.

In our implementation, the warm-start solution is generated from a minimal SA algorithm. This bitstring is interpreted as an assignment of binary values to the decision variables in the MIP model. Once verified for feasibility, the bitstring is submitted to the solver prior to the optimization phase. If accepted, it serves as the starting point for the branch-and-bound procedure, potentially establishing an incumbent solution early in the search process and reducing the number of subproblems explored. This warm-start strategy not only enhances convergence but also serves as a valuable reference for evaluating solver progress and measuring the quality of intermediate solutions throughout the optimization.

\section{Instances}
\subsection{Pareto distributed instances}\label{App:Pareto}
\begin{figure}[t]
    \centering
    \includegraphics[width=\linewidth]{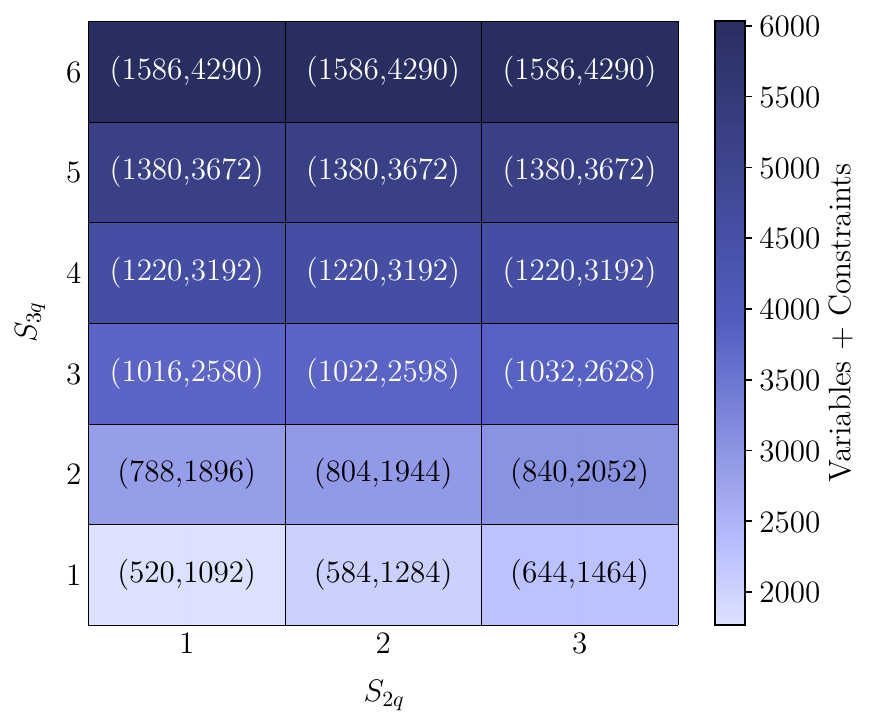}
    \caption{Pairs (variables, constraints) from CPLEX as a function of the number of sets $(S_{2q}, S_{3q})$ with $n=1$ for a $N=156$ qubits heavy-hexagonal lattice.}
    \label{fig:variables_cplex}
\end{figure}
In Figure~\ref{fig:variables_cplex}, we present the number of variables and constraints as a function of $S_{2q}$ and $S_{3q}$, which result after linearizing the HUBO problem, so it has a suitable formulation for CPLEX, for $n=1$ \textsc{swap} layer and $N=156$ qubits. Clearly, the number of variables and constraints depends on the number of sets considered, which is a consequence of the linearization of the problem. 
With the instance generator from Algorithm.~\ref{alg:swap_layers}, we can reach up to approximately $1500$ variables and $4000$ constraints using just a single \textsc{swap} layer. This reaffirms the robustness of the strategy in generating denser and complex problem instances while remaining hardware-efficient.

Additionally, we note that both the number of variables and constraints tend to saturate as the number of interaction terms increases. This behavior can be attributed to the fact that while three-body terms significantly contribute to the growth in both variables and constraints, the variables introduced by two-body terms largely constitute a subset of those already required by higher-order terms. Consequently, beyond a certain threshold, the total number of variables remains relatively stable, even as the number of interactions continues to grow.

\begin{figure}[t]
    \centering
    \includegraphics[width=1\linewidth]{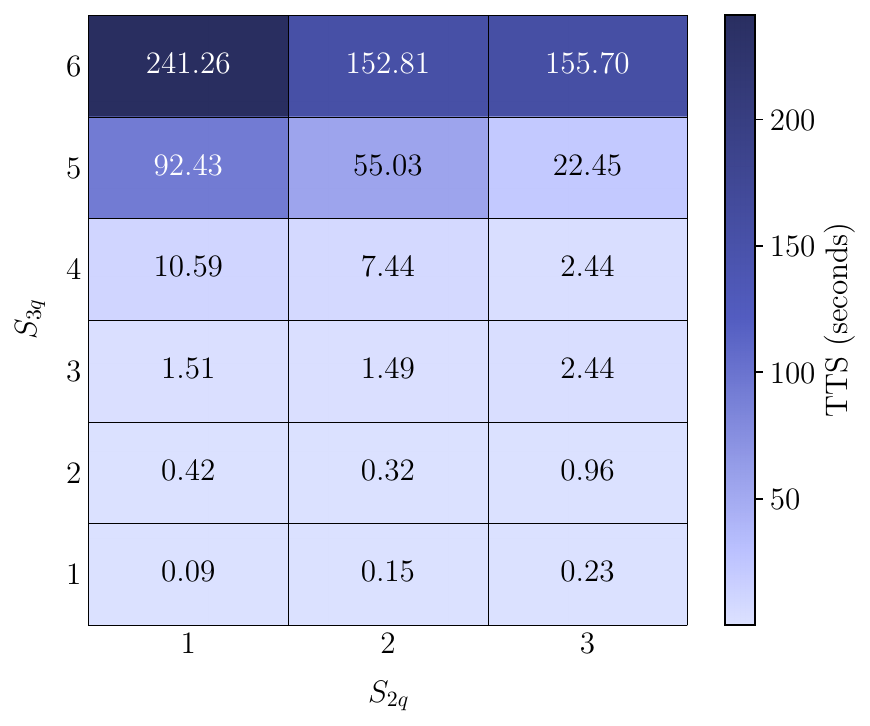}
    \caption{TTS for CPLEX as a function of the two-body and three-body terms for $N=156$ qubits $n=1$ \textsc{swap} layer and threads=1. The data shows the longest time among 250 instances drawn from a Pareto distribution with $\alpha=2$, and the coefficients were limited between -7 and 7.}
    \label{fig:max_TTS_CPLEX_250}
\end{figure}

In Figure~\ref{fig:max_TTS_CPLEX_250}, we present the maximum total time-to-solution (TTS) observed among 250 instances, where the coupling coefficients were generated using the Pareto distribution with $\alpha = 2$, and extreme values were constrained within the range $[-7, 7]$. We observe that the TTS obtained from CPLEX increases significantly from around 0.1 s to around 150 s as the number of three-body terms increases. This trend can be attributed to the fact that, in CPLEX, an increase in the number of three-body terms results in a corresponding rise in the number of variables, thereby increasing the computational complexity and making the problem more challenging to solve. 

Apart from this, the TTS follows a similar trend as observed in Figure~\ref{fig:SA_250_cauchy_156_percentcloseness}; as the number of three-body terms increases, the TTS also increases. However, for a fixed number of three-body terms, reducing the number of two-body terms leads to higher TTS values. Therefore, for our analysis comparing BF-DCQO and CPLEX, we fix $S_{2q} = 1$ and $S_{3q} = 6$. This choice ensures that experimental circuit depths with BF-DCQO remain within typical coherence times without compromising the inherent difficulty of the problem instances. It is also worthwhile to note that the runtime of the CPLEX solver depends on the specifications of the classical hardware, including factors such as clock speed, CPU architecture, number of cores, and available threads.
\begin{figure}[t]
    \centering
    \includegraphics[width=1\linewidth]{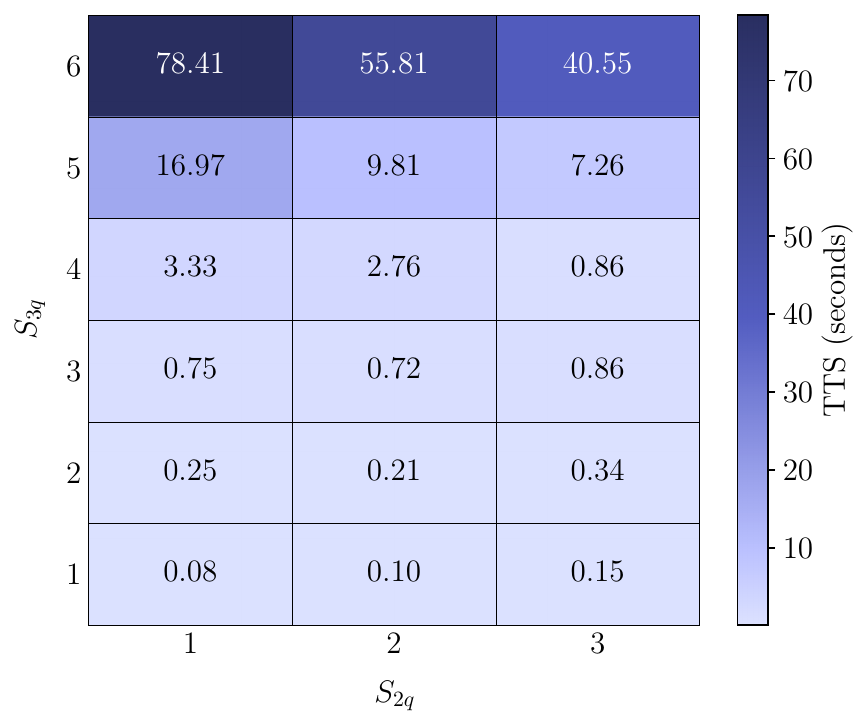}
    \caption{TTS for CPLEX as a function of the two-body and three-body terms for $N=156$ qubits $n=1$ \textsc{swap} layer and  threads=1. The data shows the mean time among 250 instances drawn from a Pareto distribution with $\alpha=2$ and the coefficients were limited between -7 and 7.}
    \label{fig:mean_TTS_CPLEX_250}
\end{figure}%

Similarly, in Figure~\ref{fig:mean_TTS_CPLEX_250}, we display the average TTS for the same set of instances. A substantial decrease from the maximum TTS values to the mean TTS values is observed, indicating that even within a distribution that poses significant challenges for CPLEX, there exists a wide variation in problem hardness across individual instances. This variation suggests that certain subsets of instances may be selectively targeted for quantum approaches, potentially increasing the likelihood of achieving a better performance.
\subsection{Cauchy distributed instances}\label{App:Cauchy}
\begin{figure}
    \centering
    \includegraphics[width=1\linewidth]{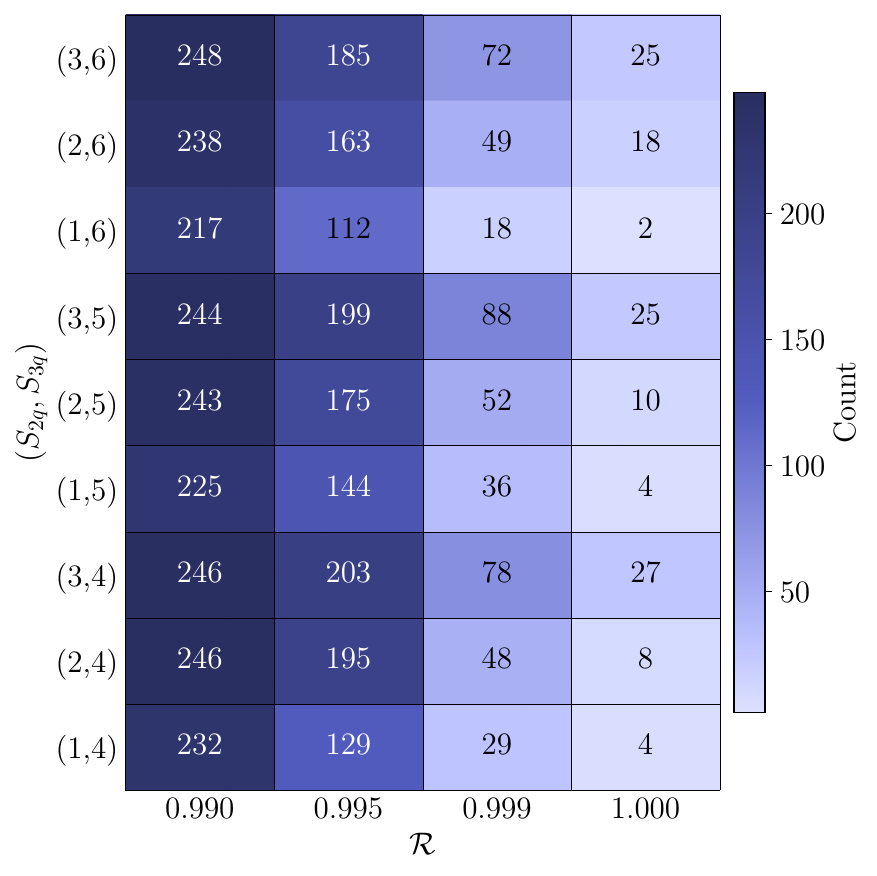}
    \caption{Number of instances where SA is within a specific $\mathcal{R}$ value. Considering $S_{2q}\in [1,3]$ and $S_{3q} \in [4,6]$ on a 156-qubit heavy-hexagonal lattice, we create $250$ random instances for each pair $(S_{2q}, S_{3q})$, with coefficients drawn from a Cauchy distribution with extreme values in the range $[-7,7]$. We show four colored tables corresponding to a $\mathcal{R}$ of (a) $0.99$, (b) $0.995$, (c) $0.999$, and (d) $1$.}
    \label{fig:SA_250_cauchy_156_percentcloseness}
\end{figure}

In Figure~\ref{fig:SA_250_cauchy_156_percentcloseness}, we consider 250 randomly generated instances and report the number of instances that achieve $\mathcal{R}\in[0.99,1]$ using $n_\text{sweep}= 20000$ and $n_\text{reads}=100$. We observe that, for a given combination of $S_{2q}$ and $S_{3q}$, a large number of instances reach at least $\mathcal{R}=0.99$; however, this number drops significantly as we approach the $\mathcal{R}=1$ threshold. This behavior suggests the presence of numerous local minima near the global optimum, making it increasingly difficult for SA to escape these traps and reach the true ground state. We observe that lower values of $S_{2q}$, corresponding to fewer two-body terms relative to three-body terms, tend to increase the problem's difficulty. This indicates that the presence of three-body interactions plays a more dominant role in determining the hardness of the instance compared to two-body terms. Additionally, once $S_{2q}$ is fixed, increasing $S_{3q}$ beyond a certain point does not significantly impact performance. 

\section{Digitized-counterdiabatic quantum computing}\label{App:dcqc}

We utilize DCQO~\cite{hegade2022digitized}, a paradigm developed to efficiently reach the ground state of a target Hamiltonian by incorporating CD protocols. In this approach, the system evolves under a time-dependent Hamiltonian given by
\begin{equation}\label{hcd}
    H_{cd}(t) = \underbrace{[1-\lambda(t)] H_m + \lambda(t) H_p}_{H_{ad}(\lambda)} + \dot{\lambda}(t) A_\lambda^{(k)},
\end{equation}
where $H_{ad}(\lambda)$ denotes the standard adiabatic Hamiltonian interpolating between the initial Hamiltonian $H_m$ and the problem Hamiltonian $H_p$ through a scheduling function $\lambda(t)$. In conventional adiabatic quantum computing, $\lambda(t)$ varies slowly to ensure that the system remains in its instantaneous ground state throughout the evolution. However, to overcome the limitations imposed by this slow-driving requirement, DCQO introduces a velocity-dependent counterdiabatic term $\dot{\lambda}(t) A_\lambda^k$, where $ A_\lambda^{(k)}$ represents a $k$-order approximation to the adiabatic gauge potential (AGP).

The approximate AGP takes the form $ A_\lambda^{(k)} = i \sum_k \beta_k O_{2k-1}$, where each operator $O_k$ is recursively defined as $O_k = \mathcal{L}^k \partial_\lambda H_{ad}(\lambda)$, with $\mathcal{L}(\circ) = [H_{ad}, \circ]$ denoting the Liouvillian super-operator. The coefficients $\beta_k(t)$ can be determined using methods such as action minimization~\cite{doi:10.1073/pnas.1619826114,PhysRevLett.123.090602} or Krylov subspace techniques~\cite{PhysRevX.14.011032,bhattacharjee2023lanczos}. 

Starting from the ground state of $H_m$, the system can undergo a digitized evolution composed of $n_\text{trot}$ Trotter steps. This evolution is approximated by a product of unitary operators of the form $U(j) =  \exp\left\{-i H_{cd}(j \Delta t)\, \Delta t\right\}$, where $\Delta t$ is the step size. The full time evolution from $t=0$ to $t=T$ is then constructed as $U(0,T) = \prod_{j=1}^{n_{\text{trot}}} U(j)$, where $n_{\text{trot}}$ denotes the total number of Trotter steps. Due to the presence of the counterdiabatic term $A_\lambda^{(k)}$ in the Hamiltonian $H_{cd}(t)$, this digitized evolution enables an evolution compared to standard adiabatic evolution. As a result, the system can reach the ground state of $H_p$ more efficiently, with significantly reduced circuit depths, making the protocol more suitable for implementation on quantum hardware~\cite{hegade2022digitized,YaoLinBukov21}. Recently, it was demonstrated that circuit depth can be further reduced in the impulse regime, where $|\alpha_k(t) \dot{\lambda}(t)| \gg |\lambda(t)|$~\cite{PhysRevApplied.20.014024,cadavid2023efficient}. In this limit, the contributions from $H_{ad}(\lambda)$ can be neglected, as the CD term predominantly governs the system's dynamics~\cite{Carolan22}.
\begin{table*}[htb]
    \centering
    \caption{Experiments: Performance analysis of BF-DCQO and SA with $n=1$, $S_{2q}=1$, $S_{3q}=4$. ``SA 10000 (1000)" shows minimum energy obtained with 100 runs of SA. $\SI{0.6e-5}{s}$ per sweep CPU time and $10^4$ shots per sec QPU time.}
    \label{tab:EXP_SA_combined}
    \begin{ruledtabular}\begin{tabular}{cccccccccc}
        $N$ & Instance & $n_{\rm{iter}}$ & Optimal cost & SA 10000 & SA 1000 & BF-DCQO & CPU time & QPU time & Total time \\
        \midrule
        100 & 0  & 0 & -109.1738 & -109.1738 & -107.2830 & \textbf{-107.8784} & 0.301 & 0.2 & 0.501 \\
        100 & 1 & 0 & -103.9619 & -102.7143 & -101.0449 & -99.0469          & 0.301 & 0.5 & 0.801 \\
        100 & 2 & 0 & -110.8313 & -109.9482 & -109.3494 & -109.3201         & 0.301 & 0.5 & 0.801 \\
        100 & 3 & 0 & -89.1511  & -87.8355  & -86.8249  & -86.4880          & 0.801 & 0.2 & 0.801 \\
        \midrule
        130 & 0 & 2 & -115.9914 & -115.3305 & -114.1877 & -115.1009         & 0.308 & 6.0 & 6.308 \\
        130 & 1  & 2 & -161.4084 & -161.4084 & -158.9392 & -159.5753         & 0.318 & 1.2 & 1.518 \\
        130 & 2 & 2 & -129.5597 & -128.4072 & -128.1709 & -127.7192         & 0.318 & 1.2 & 1.518 \\
        130 & 3 & 2 & -140.5795 & -140.4566 & -138.6439 & -139.0727         & 0.318 & 1.2 & 1.518 \\
        130 & 4  & 3 & -135.6783 & -134.6596 & -134.0120 & \textbf{-135.1691}& 0.174 & 2.2 & 2.374 \\
        \midrule
        156 & 0 & 1 & -191.1775 & -188.3061 & -183.9224 & \textbf{-188.5845}& 0.612 & 0.7 & 1.311 \\
        156 & 1 & 1 & -157.9815 & -156.0199 & -153.6049 & -154.4282         & 0.072 & 0.7 & 0.772 \\
        156 & 2 & 1 & -182.1461 & -178.8977 & -178.4260 & -178.1232         & 0.072 & 1.2 & 1.272 \\
        156 & 3  & 1 & -165.6203 & -164.5468 & -162.2066 & -163.2233         & 0.132 & 1.2 & 1.331 \\
        156 & 4  & 2 & -161.2242 & -159.4274 & -158.2813 & -157.6393         & 0.318 & 1.2 & 1.518 \\
    \end{tabular}\end{ruledtabular}
\end{table*}
\begin{table*}[htb]
    \centering
    \caption{Experiments: Performance analysis of BF-DCQO and CPLEX with $n=1$, $S_{2q}=1$, $S_{3q}=6$. ``TT$_\mathcal{R}$ (BF-DCQO)'' is total BF-DCQO wall time (CPU + QPU). ``TT$_\mathcal{R}$ (CPLEX)'' is time-to-solution for CPLEX. $\SI{0.6e-5}{s}$ per sweep CPU time and $10^4$ shots per sec QPU time.}
    \label{tab:EXP_CPLEX_combined}
    \begin{ruledtabular}\begin{tabular}{ccccccccccc}
        $N$ & Instance & $n_{\rm{iter}}$ & Optimal cost & TTS (CPLEX) & BF-DCQO & CPU time & QPU time & TT$_\mathcal{R}$ (BF-DCQO) & TT$_\mathcal{R}$ (CPLEX) & Enhancement Factor \\
        \midrule
        80  & 0  & 2 & -215.4481 & 16.1915 & \textbf{-201.6557}  & 0.0180 & 1.2  & 1.218  & 4.5  & 3.694 \\
        80  & 1  & 1 & -222.2756 & 15.0387 & \textbf{-207.7738}  & 0.0120 & 0.7  & 0.712  & 4.5  & 6.318 \\
        80  & 2 & 1 & -216.3627 & 13.7088 & \textbf{-202.9397}  & 0.0120 & 0.7  & 0.712  & 4.5  & 6.318 \\
        80  & 3  & 3 & -227.9471 & 12.1781 & \textbf{-221.7880}  & 0.0241 & 2.2  & 2.224  & 7.5  & 3.372 \\
        80  & 4 & 3 & -230.1366 & 12.0960 & \textbf{-226.2100}  & 0.0241 & 2.2  & 2.224  & 4.5  & 2.023 \\
        \midrule
        100 & 0 & 2 & -265.7944 & 57.9967 & \textbf{-257.3076}  & 0.0186 & 1.2  & 1.219  & 9.0  & 7.387 \\
        100 & 1 & 2 & -282.9621 & 52.8523 & \textbf{-273.7422}  & 0.0186 & 1.2  & 1.219  & 6.5  & 5.334 \\
        100 & 2 & 0 & -281.6533 & 51.4635 & \textbf{-271.1121}  & 0.0060 & 0.2  & 0.207  & 10.0 & 48.396 \\
        100 & 3  & 2 & -253.2525 & 46.6339 & \textbf{-244.1057}  & 0.0186 & 1.2  & 1.219  & 6.5  & 5.335 \\
        100 & 4 & 3 & -266.9979 & 46.2196 & \textbf{-258.0769}  & 0.0246 & 2.2  & 2.225  & 6.0  & 2.697 \\
        \midrule
        130 & 0  & 2 & -364.1344 & 130.0527 & \textbf{-344.8771} & 0.0186 & 1.2  & 1.218  & 12.0 & 9.852 \\
        130 & 1 & 1 & -355.1786 & 100.1148 & \textbf{-343.9632} & 0.0126 & 0.7  & 0.712  & 12.5 & 17.556 \\
        130 & 2 & 3 & -379.2107 & 99.2038  & \textbf{-370.7371} & 0.0246 & 2.2  & 2.224  & 17.5 & 7.867 \\
        130 & 3 & 3 & -377.8902 & 97.1254  & \textbf{-345.3099} & 0.0126 & 2.2  & 2.224  & 4.5  & 2.023 \\
        130 & 4  & 1 & -368.7392 & 103.3247 & \textbf{-354.8130} & 0.0126 & 0.7  & 0.712  & 15.5 & 21.765 \\
        \midrule
        156 & 0 & 0 & -472.5398 & 241.25   & \textbf{-454.0458} & 0.0066 & 0.2  & 0.207  & 17.5 & 84.729 \\
        156 & 1  & 3 & -465.7531 & 198.42   & \textbf{-425.5693} & 0.0246 & 1.7  & 1.725  & 8.0  & 4.639 \\
        156 & 2 & 3 & -464.0377 & 125.67   & \textbf{-443.7394} & 0.0246 & 1.7  & 1.725  & 8.5  & 4.930 \\
        156 & 3 & 2 & -466.9106 & 122.37   & \textbf{-444.0443} & 0.0186 & 1.2  & 1.219  & 19.0 & 15.595 \\
        156 & 4 & 1 & -439.4456 & 123.34   & \textbf{-406.6555} & 0.0126 & 0.7  & 0.713  & 16.0 & 22.448 \\
    \end{tabular}\end{ruledtabular}
\end{table*}
\begin{table*}[htb]
    \centering
    \caption{Experiments: Performance analysis of BF-DCQO and CPLEX for varying $N$ values with $n=1$, $S_{2q}=1$, $S_{3q}=6$, $\SI{0.6e-5}{s}$ per sweep CPU time and $10^4$ shots per sec QPU time.}
    \begin{ruledtabular}\begin{tabular}{lccccc}
        $N$ & Instance & TTS (CPLEX) & TT$_\mathcal{R}$ (BF-DCQO) & TT$_\mathcal{R}$ (CPLEX) & TT$_\mathcal{R}$ (SA + CPLEX) \\
        \midrule
            & 0  & 16.19 & 1.218  & 4.5  & 5.5 \\
            & 1  & 15.03 & 0.712  & 4.5  & 4.5 \\
        80  & 2 & 13.70& 0.712  & 4.5  & 4.0 \\
            & 3  & 12.17 & 2.224  & 7.5  & 6.0 \\
            & 4 & 12.09 & 2.224  & 4.5  & 7.5 \\
        \midrule
            & 0 & 57.99 & 1.2186 & 9.0  & 7.0 \\
            & 1 & 52.85 & 1.2186 & 6.5  & 8.0 \\
        100 & 2 & 51.46 & 0.2066 & 10.0 & 1.0 \\
            & 3  & 46.63 & 1.2186 & 6.5  & 7.5 \\
            & 4 & 46.21 & 2.2246 & 6.0  & 10.0 \\
        \midrule
            & 0  & 130.05 & 1.218  & 12.0 & 16.5 \\
            & 1 & 100.11 & 0.712  & 12.5 & 6.5 \\
        130 & 2 & 99.20  & 2.224  & 17.5 & 14.0 \\
            & 3 & 97.12  & 2.224  & 4.5  & 0.5 \\
            & 4  & 103.32 & 0.712  & 15.5 & 15.5 \\
        \midrule
            & 0 & 241.25     & 0.2066 & 17.5 & 14.0 \\
            & 1  & 198.42     & 1.7246 & 8.0  & 13.5 \\
        156 & 2 & 125.67     & 1.7246 & 8.5  & 9.5 \\
            & 3 & 122.37     & 1.2186 & 19.0 & 9.5 \\
            & 4 & 123.34     & 0.7125 & 16.0 & 17.0 \\
    \end{tabular}\end{ruledtabular}
    \label{tab:EXP_CPLEX_all_N}
\end{table*}

\section{Runtime calculations}\label{App:runtimes}
\begin{figure}
    \centering
    \includegraphics[width=1\linewidth]{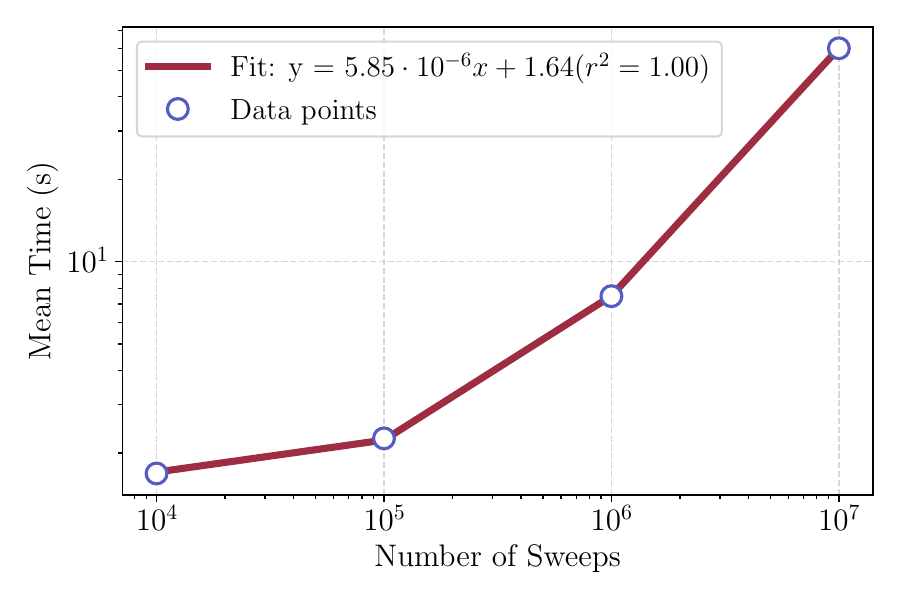}
    \caption{Mean times as a function of number of sweeps for SA for a random $N=156$ qubit instance with $S_{2q} = 1$ and $S_{3q}=6$. The plot depicts mean times of 100 runs with $10^4$, $10^5$, $10^6$, and $10^7$ sweeps.  }
    \label{fig:Time_estimation_per_sweep}
\end{figure}
Before analyzing the performance of SA, we first estimate the time per sweep $T_{\rm{sweep}}$, using the CPU specifications listed in Table~\ref{tab:specs}. To this end, we executed SA for a random $N=156$ qubit instance, with an increasing number of sweeps ranging from $10^4$ to $10^7$, with 100 runs each. The average runtime $T_{\rm{avg}}$ required to obtain the output was recorded and is presented in Figure~\ref{fig:Time_estimation_per_sweep}. 

A linear fit of the form $T_{\rm{avg}} = T_{\rm{sweep}}~ n_{\rm{sweep}} + T_{\rm{offset}}$ was performed, yielding a time per sweep of $T_{\rm{sweep}} \approx \SI{0.6e-5}{s}$. Additionally, we observed an offset time of $T_{\rm{offset}} \approx \SI{1.65}{s}$, which can be attributed to initialization overheads such as solver setup and memory allocation. Henceforth, we fix the sweep time to $T_{\rm{sweep}}$ for calculating the CPU runtimes associated with SA, as well as for estimating the classical pre- and post-processing time in the BF-DCQO pipeline. 

\section{Experimental results}\label{App:additional_SA}

\emph{Outperforming SA.}---In Table~\ref{tab:EXP_SA_combined} we compare the performance of BF-DCQO against SA using instances generated with a single \textsc{swap} layer ($n=1$), $S_{2q}=1$, and $S_{3q}=4$. The coupling coefficients were sampled from a Cauchy distribution, as described previously. Hard instances were selected based on the energy gap between the optimal cost and the minimum energy obtained via SA with $n_\text{sweep} = 20000$ and $n_\text{runs}=100$. As discussed earlier, SA tends to reach solutions that are approximately $99\%$ close to optimal relatively quickly, but requires significantly more time to converge to the exact solution. 

\emph{Outperforming CPLEX.}---The experimental results comparing CPLEX and BF-DCQO are presented in Table~\ref{tab:EXP_CPLEX_combined}. For $N=80$ qubits, all five instances, the initial SA pre-processing step was configured $n_\text{sweep}^\text{pre}=10$ sweeps and a single run, given that the average runtime for CPLEX is approximately $13$ seconds. Here, $n_{\rm{iter}} = 0$ corresponds to SA-initialized DCQO without any bias field updates, while $n_{\rm{iter}} = 1,2,\dots$ denote subsequent iterations with updated bias fields. All reported times are in seconds. The column labeled ``Optimal Cost'' indicates the minimum energy obtained by CPLEX, with the corresponding provable optimality runtime shown in the column ``TTS (CPLEX)''. The column ``BF-DCQO'' reports the minimum energy achieved using our proposed algorithm, along with the associated CPU and QPU times. Additionally, ``TT$_\mathcal{R}$ (CPLEX)'' represents the time taken by CPLEX to reach the same energy level as that obtained by BF-DCQO (or lower), where CPLEX solutions are sampled every 0.5 seconds. Enhancement factor denotes the ratio between the TT$_\mathcal{R}$ from CPLEX and BF-DCQO.

It can be observed that for all considered instances with $N=80$ qubits, TT$_\mathcal{R}$ for BF-DCQO is substantially lower than that of CPLEX. In particular, for some instances, BF-DCQO achieves approximately a six-fold reduction in total runtime compared to CPLEX. It is important to note that, in these runtime calculations, we exclude the Qiskit transpilation times. Furthermore, since CPLEX is a deterministic exact solver and BF-DCQO is inherently probabilistic, we compare only the minimum energy obtained by each method within a given runtime window to ensure a fair comparison.

For $N=100$, $N=130$, and $N=156$ qubits, these instances, the SA pre-processing step was configured with 100 sweeps and a single run, consistent with the earlier setup. We observe performance trends similar to the $N=80$ case, with BF-DCQO consistently outperforming CPLEX in terms of TT$_\mathcal{R}$. Notably, for certain instances, we find that even at $n_{\rm{iter}}=0$, the algorithm yields competitive results. This highlights the robustness of the BF-DCQO approach, further enhanced through the integration of classical pre- and post-processing steps. 

In Table~\ref{tab:EXP_CPLEX_all_N}, we observe that even when CPLEX is warm-started with the same bitstring, BF-DCQO continues to demonstrate a comparable performance advantage in most cases. Interestingly, for certain instances, the warm-started CPLEX exhibits even longer runtimes compared to standard CPLEX runs without initialization. This behavior indicates that the quantum component of BF-DCQO plays a dominant role in driving the observed performance improvements. However, in one specific instance for $N=130$ qubits, the warm-started CPLEX outperforms BF-DCQO. Nevertheless, on average, BF-DCQO consistently outperforms CPLEX across all instances considered, reinforcing its effectiveness in solving challenging HUBO problems.

\bibliography{reference.bib}
\clearpage

\end{document}